\documentstyle[epsf,rotate]{aipproc}

\begin{document}
\pagestyle{plain}

\title{Algebraic model of baryon resonances}

\author{R. Bijker$^{1)}$ and A. Leviatan$^{2)}$}
\address{
$^{1)}$ Instituto de Ciencias Nucleares, 
        Universidad Nacional Aut\'onoma de M\'exico, 
        A.P. 70-543, 04510 M\'exico D.F., M\'exico\\
$^{2)}$ Racah Institute of Physics, 
        The Hebrew University, Jerusalem 91904, Israel}

\maketitle

\begin{abstract}
We discuss recent calculations of electromagnetic form factors 
and strong decay widths of nucleon and delta resonances. The  
calculations are done in a collective constituent model of the nucleon, 
in which the baryons are interpreted as rotations and vibrations
of an oblate top. 
\end{abstract}

\section{Introduction}

The study of the properties of baryon resonances is entering 
a new era with the forthcoming new and more accurate data from 
new facilities, such as Jefferson Lab., MAMI, ELSA and Brookhaven. 
Effective models of baryons which are based on three constituents 
($qqq$) share a common spin-flavor-color structure but differ in their  
assumptions on the spatial dynamics. For example, 
quark potential models in nonrelativistic \cite{NRQM} or relativized 
\cite{RQM} forms emphasize the single-particle aspects of quark dynamics  
for which only a few low-lying configurations in the confining potential 
contribute significantly to the eigenstates of the Hamiltonian. 
On the other hand, some regularities 
in the observed spectra, such as linear Regge trajectories and 
parity doubling, hint that an alternative, collective 
type of dynamics may play a role in the structure of baryons. 

In this contribution we discuss a collective model within the 
context of an algebraic approach \cite{BIL} and present some 
results for electromagnetic form factors \cite{emff} and strong 
decay widths \cite{strong}.

\section{Collective model of baryons} 

\begin{figure}
\centering
\epsfxsize 1.75in
\rotate{\rotate{\rotate{\epsffile{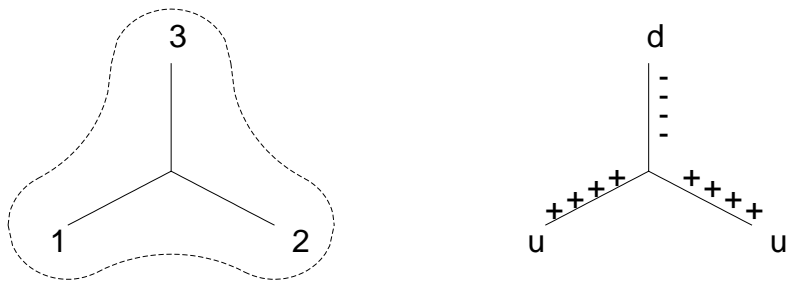}}}}
\caption[]{
Collective model of baryons (the charge
distribution of the proton is shown as an example).}
\label{rb:geometry}
\end{figure} 

We consider a collective model in which the baryon resonances
are interpreted in terms of rotations and vibrations of the string 
configuration in Fig.~\ref{rb:geometry}. 
A fit to the 3 and 4 star nucleon and delta resonances 
gives a r.m.s. deviation of 39 MeV \cite{BIL}.
The corresponding oblate top wave functions are spread over many 
oscillator shells and hence are truely collective. The baryon wave 
functions have the form
\begin{eqnarray}
\left| \, ^{2S+1}\mbox{dim}\{SU_f(3)\}_J \,
[\mbox{dim}\{SU_{sf}(6)\},L^P]_{(v_1,v_2);K} \, \right> ~. \label{wf}
\end{eqnarray}
The spin-flavor part has the usual $SU_{sf}(6)$ classification and
determines the permutation symmetry of the state.
The spatial part is 
characterized by the labels: $(v_1,v_2);K,L^P$, where $(v_1,v_2)$
denotes the vibrations (stretching and bending) of the string configuration
in Fig.~\ref{rb:geometry}; $K$ denotes the projection of the
rotational angular momentum $L$ on the body-fixed symmetry-axis and 
$P$ the parity. Finally, $S$ and $J$ are the spin and the total 
angular momentum $\vec{J}=\vec{L}+\vec{S}$~.
In this contribution we focus on the nucleon resonances 
which are interpreted as rotational excitations of the 
$(v_1,v_2)=(0,0)$ vibrational ground state. 

The electromagnetic (strong) coupling is assumed to 
involve the absorption or emission of a photon (elementary meson) 
from a single constituent. 
The collective form factors and decay widths are obtained by 
folding with a probability distribution for the charge and 
magnetization along the strings of Fig.~\ref{rb:geometry}
\begin{eqnarray}
g(\beta) &=& \beta^2 \, \mbox{e}^{-\beta/a}/2a^3 ~, \label{gbeta}
\end{eqnarray}
where $\beta$ is a radial coordinate and $a$ is a scale parameter.
In the algebraic approach these form factors and decay widths 
can be obtained in closed analytic form (in the limit of a large 
model space) which allows us to do a straightforward
and systematic analysis of the experimental data.

The ansatz of Eq.~(\ref{gbeta}) for the probability distribution
is made to obtain a dipole form for the proton electric form factor
\begin{eqnarray}
G_E^p(k) &=& \frac{1}{(1+k^2a^2)^2} ~. 
\end{eqnarray}
The same distribution is used to calculate transition form factors 
and decay widths. As a result, all collective form factors are found 
to drop as powers of the momentum transfer \cite{emff}. 
This property is well-known 
experimentally and is in contrast with harmonic oscillator based quark 
models in which all form factors fall off exponentially.

\section{Electroproduction}

Electromagnetic inelastic form factors can be measured in 
electroproduction 
of baryon resonances. They are expressed in terms of helicity amplitudes 
$A_{\nu}^N$, where $\nu$ indicates the helicity and $N$ 
represents proton ($p$) or neutron ($n$) couplings.
In a string-like model of hadrons one expects \cite{johnsbars} 
on the basis of QCD that strings will elongate (hadrons swell)
as their energy increases. This effect can be
easily included in the present analysis by making the scale parameters
of the strings energy- dependent. We use here the simple ansatz
\begin{eqnarray}
a &=& a_0\Bigl ( 1 + \xi\,{W-M\over M}\Bigr ) ~, \label{stretch}
\end{eqnarray}
where $M$ is the nucleon mass and $W$ the resonance mass. This ansatz
introduces a new parameter ($\xi$), the stretchability of the string.

\begin{figure}[b]
\epsfxsize 5.5cm
\hspace{3cm}\rotate{\rotate{\rotate{\epsffile{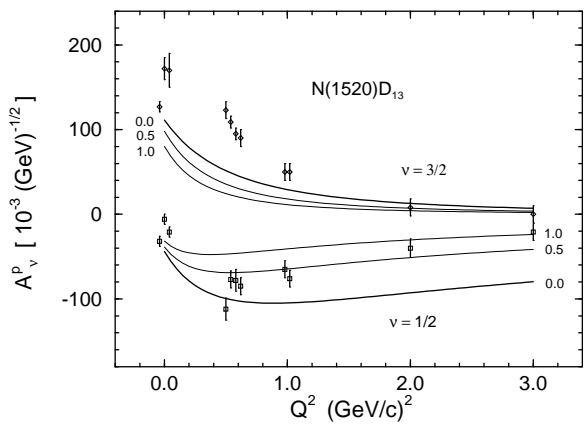}}}}
\vspace{0.5cm}
\caption[]{Effect of hadron swelling for excitation of $N(1520)D_{13}$. 
The curves are labelled by the stretching parameter $\xi$
of Eq.~(\ref{stretch}).}
\label{rb:d13}
\end{figure}

In Fig.~\ref{rb:d13} we show the transverse helicity amplitudes 
$A^{p}_{\nu}$ for the $N(1520)D_{13}$ resonance calculated in the Breit 
frame (a factor of $+i$ is suppressed). The scale parameter $a$ is 
determined in a simultaneous fit to the nucleon charge radii and the  
nucleon elastic form factors: $a=0.232$ fm \cite{emff}.  
The effect of stretching on the helicity amplitudes is sizeable 
(especially if one takes the value $\xi\approx 1$ which is suggested
by QCD arguments \cite{johnsbars} and the Regge behavior of nucleon
resonances). The data for $N(1520)D_{13}$ (and also for $N(1680)F_{15}$) 
show a clear indication that the form factors are dropping faster than 
expected on the basis of the dipole form. 
The disagreement at low momentum transfer may be due to the neglect 
of the meson cloud. 

\section{Strong decay widths}

\begin{table}[b]
\centering
\caption[]{$N \pi$ and $N \eta$ decay widths of 3 and 4 star nucleon
resonances in MeV.}
\vspace{0.5cm}
\label{rb:strong}
\begin{tabular}{lcccccc}
& & & & & & \\
State & Mass & Resonance & \multicolumn{2}{c}{$\Gamma(N\pi)$} 
& \multicolumn{2}{c}{$\Gamma(N\eta)$} \\
& & & th & exp & th & exp \\
& & & & & & \\
\hline
& & & & & & \\
$S_{11}$ & $N(1535)$ & $^{2}8_{1/2}[70,1^-]_{(0,0);1}$
&  $85$ & $79 \pm 38$ & 0.1 & $74 \pm 39$ \\
$S_{11}$ & $N(1650)$ & $^{4}8_{1/2}[70,1^-]_{(0,0);1}$
&  $35$ & $130 \pm 27$ & 8 & $11 \pm 6$ \\
$P_{13}$ & $N(1720)$ & $^{2}8_{3/2}[56,2^+]_{(0,0);0}$
&  $31$ & $22 \pm 11$ & 0.2 & \\
$D_{13}$ & $N(1520)$ & $^{2}8_{3/2}[70,1^-]_{(0,0);1}$
& $115$ & $67 \pm 9$ & 0.6 & \\
$D_{13}$ & $N(1700)$ & $^{4}8_{3/2}[70,1^-]_{(0,0);1}$
&   $5$ & $10 \pm 7$ & 4 & \\
$D_{15}$ & $N(1675)$ & $^{4}8_{5/2}[70,1^-]_{(0,0);1}$
&  $31$ & $72 \pm 12$ & 17 & \\
$F_{15}$ & $N(1680)$ & $^{2}8_{5/2}[56,2^+]_{(0,0);0}$
&  $41$ & $84 \pm 9$ & 0.5 & \\
$G_{17}$ & $N(2190)$ & $^{2}8_{7/2}[70,3^-]_{(0,0);1}$
&  $34$ & $67 \pm 27$ & 11 & \\
$G_{19}$ & $N(2250)$ & $^{4}8_{9/2}[70,3^-]_{(0,0);1}$
&  $7$ & $38 \pm 21$ & 9 & \\
$H_{19}$ & $N(2220)$ & $^{2}8_{9/2}[56,4^+]_{(0,0);0}$
&  $15$ & $65 \pm 28$ & 0.7 & \\
$I_{1,11}$ & $N(2600)$ & $^{2}8_{11/2}[70,5^-]_{(0,0);1}$
&  $9$ & $49 \pm 20$ & 3 & \\
& & & & & & \\
\hline
\end{tabular}
\end{table}

In addition to electromagnetic couplings, strong decays provide an 
important, complementary, tool to study the structure of baryons.
We consider decays with emission of $\pi$ and $\eta$.
The experimental data \cite{PDG} are shown in Table~\ref{rb:strong}, 
where they are compared with the results of our calculation. 
The calculated values depend on two parameters $g$ and $h$ in the 
transition operator and on the scale parameter $a$ of Eq.~(\ref{gbeta}).
These parameters are determined from a least-square fit to the $N \pi$ 
partial widths (which are relatively well known) with the exclusion
of the $S_{11}$ resonances whose assignments are not clear due to 
possible mixing between the $N(1535)$ and $N(1650)$ and/or 
the possible presence of a third $S_{11}$ resonance \cite{LW}.

The calculation of $N \pi$ decay widths is found to be in fair 
agreement with experiment (see Table~\ref{rb:strong}). 
These results are to a large extent a consequence of spin-flavor symmetry. 
The calculated widths for the $N \eta$ channel are systematically 
small. We emphasize here that, since the transition operator was determined
from the $N \pi$ decays, the $\eta$ decays are
calculated without introducing any further parameters.
The results of this analysis suggest that the large $\eta$ width for the
$N(1535)S_{11}$ is not due to a conventional $q^3$ state. One possible
explanation is the presence of another state in the same mass region,
{\it e.g.} a quasi-bound meson-baryon $S$ wave resonance just below
or above threshold, for example $N\eta$, $K\Sigma$ or $K\Lambda$
\cite{Kaiser}. Another possibility is an exotic configuration of four
quarks and one antiquark ($q^{4}\bar{q}$).

\section{Summary and conclusions}

We have analyzed simultaneously electromagnetic form factors and 
strong decay widths in a collective model of the nucleon. The helicity 
amplitudes are folded with a probability distribution for the charge 
and magnetization, which is determined from the well-established 
dipole form of the nucleon electromagnetic form factors. 
The same distribution function is used to calculate the transition 
form factors for the excited baryons. As a result, all form factors 
drop as powers of the momentum transfer. 

For electromagnetic couplings we find that the inclusion of the 
stretching of baryons improves the calculation of the helicity amplitudes 
for large values of the momentum transfer. 
The disagreement for small values of the momentum transfer 
$0 \leq Q^2 \leq 1$ (GeV/c)$^2$ may be due to coupling of the photon
to the meson cloud, {\it i.e.} configurations of the type
$q^3-q\bar{q}$. Since such configurations have much larger spatial 
extent than $q^3$, their effects are expected to drop faster with 
momentum transfer than the constituent form factors.

An analysis of the strong decay widths into the $N \pi$ and $N \eta$ 
channels shows that the $\pi$ decays follow the expected pattern.
Our calculations do not show any indication for a large $\eta$ width,
as is observed for the $N(1535)S_{11}$ resonance. The observed large
$\eta$ width indicates the presence of another configuration, which is
outside the present model space. 

The results reported in this article are based on work done
in collaboration with F. Iachello (Yale).
The work is supported in part by grant No. 94-00059 from the United 
States-Israel Binational Science Foundation (BSF), Jerusalem, Israel 
(A.L.) and by CONACyT, M\'exico under project 
400340-5-3401E and DGAPA-UNAM under project IN105194 (R.B.).

\end{document}